\documentclass[11pt, letterpaper]{article}
\usepackage{amsfonts,amsmath,amsopn,amssymb,amsthm,bbm,latexsym,mathrsfs,pstricks,verbatim,color}

\font\small=cmr8

\font\tenmsy=msbm10
\font\sevenmsy=msbm10 at 7pt
\font\fivemsy=msbm10 at 5pt
\newfam\msyfam 
\textfont\msyfam=\tenmsy
\scriptfont\msyfam=\sevenmsy
\scriptscriptfont\msyfam=\fivemsy

\def\z{{\cal Z}}

\def\p{{\tilde p}}

\def\y{{\infty}}

\def\rw{\rightarrow}
\def\lrw{\leftrightarrow}

\let\rw\rightarrow

\def\M{{\cal M}}

\font\small=cmr8

\begin{document}

\vskip18pt

\title{\vskip60pt {\bf Nonlocal operator basis from the path representation of the 
$\M(k+1,k+2)$ and the $\M(k+1,2k+3)$  minimal  models }}

\vskip18pt


\smallskip
\author{ \bf{P. Jacob$^{1,2}$ and P.
Mathieu$^2$}\thanks{patrick.jacob@phy.ulaval.ca,
pmathieu@phy.ulaval.ca.  This work is supported by  NSERC.} \\ 
\\
${}^1$Department of Mathematical Sciences, \\University of Durham, Durham, DH1 3LE, UK\\
and\\
${}^2$D\'epartement de physique, de g\'enie physique et d'optique,\\
Universit\'e Laval,
Qu\'ebec, Canada, G1K 7P4.
}

\vskip .2in
\bigskip

\maketitle


\vskip0.3cm
\centerline{{\bf ABSTRACT}}
\vskip18pt
We reinterpret  a path describing a state in an irreducible module of the unitary minimal model ${\cal M} (k+1,k+2)$ in terms of a string of charged operators acting on the module's ground-state path. Each such operator acts non-locally on a path. The path characteristics are then translated into a set of conditions on sequences of operators that provide an operator basis. As an application, we re-derive  the vacuum finite fermionic character by constructing the generating function of these basis states.
 These results generalize directly to  the  ${\cal M} (k+1,2k+3)$ models, the
close relatives of the unitary models in terms of path description.



\section{Introduction}

\subsection{General orientation}

The minimal models {\cite{BPZ} are solvable models that describe universality classes of two-dimensional critical phenomena. As such, they are physically quite important. 
Their exact solvability is rooted in the representation theory of the Virasoro algebra. Every field is in correspondence with a state and the space of states is a finite direct sum of highest-weight modules. In each such module, there is an infinite number of singular vectors. These leads to differential equations  for the correlation functions. This accounts for the models exact solvability. Irreducible modules are obtained by factoring out these singular vectors using an inclusion-exclusion process that results into an expression of the character in the form of an  infinite alternating sum (see e.g., \cite{CFT}). Although these characters can be obtained in closed forms, the states in the irreducible modules do not have a clear physical interpretation in this description.

The discovery of fermionic   character formulae \cite{KKMMa,KKMMb}, which are positive multiple sums, suggests the existence of an alternative and more physical approach to the representation theory of the Virasoro algebra that would be based on the concept of quasi-particles. In this picture, the Hilbert space is constructed by a filling process subject to restriction rules. Along that line, a useful guide is provided by the realization of  the minimal models as statistical models, in particular, the restricted-solid-on-solid (RSOS) models of \cite{ABF,FB}. The solution of these models embody a description of the states in terms of  configuration sums (or paths) that are tailor-made for a reinterpretation in terms of quasi-particles. Once the quasi-particles are identified, the next step would be to interpret them in a conformal-field-theoretical setting.

The present work lies mid-way within this program. Roughly, we abstract the path description of a particular class of minimal models 
 into an operator basis whose interpretation is yet to be formulated. This is clarified in the following subsections.

\subsection{Unitary minimal models: paths and heuristic considerations underlying their operator interpretation}

For the larger part of this work, we will be concerned with the unitary minimal models ${\cal M} (k+1,k+2)$ \cite{FQS}. Their
 states have a particularly simple representation as lattice paths. 
 It is inherited from the models off-critical formulation in terms of the Andrews-Baxter-Forrester 
 RSOS models in regime III \cite{ABF, Huse}.
More precisely, the solution of this statistical model  by  the corner-transfer-matrix method induces a representation of every state in the form of a particular configuration. A configuration is the specification of the set of values of the order parameter, the height variable, say $y$,  for all integral positions $x$ between 0 and a certain length  $L$. The height is bounded by the parameter $k$ characterizing the model:  $0\leq y\leq k$ (where $k$ is related to the parameter $r$ of \cite{ABF} by $r=k+2$). Every configuration  is uniquely specified by its contour obtained by  linking  adjacent heights.  Since two adjacent heights can differ by $\pm1$, the links are either  North-East (NE) or South-East (SE) edges. This contour defines an integer-lattice path.

 
 To each path, we associate a weight. The weight of a path is simply the sum of the weight of all its vertices in between 1 and $L-1$; each  vertex has weight $x/2$, where $x$ is its horizontal position, unless it is a local maximum or minimum, in which case its weight is  0.

Classes of paths are specified by the boundary conditions, namely the pairs of values $(y_0,y_L)$.
Quite remarkably, the ($q$-weighted) generating function for all paths with fixed boundary conditions becomes, in the limit where $L\rw\y$, is equal to the (normalized) character of the irreducible module $\chi_{r,s}(q)$, for some pair $(r,s)$  determined by $(y_0,y_L)$, of the ${\cal M} (k+1,k+2)$ models \cite{Kyoto}.
Each path is thus a representation of a state in a specific unitary  irreducible module and a finite path 
describes a state in a natural finitization of the module  \cite{Mel}.

Characters can be read-off this path representation in at least two different ways. We can derive recurrence relations on the finite-path generating function by considering the effect of removing an edge at the end of the path \cite{ABF}. The immediate solution of these relations  yields a finitized version of the Rocha-Caridi (or bosonic) character formulae. A different approach consists in considering the generating function for paths as the grand-canonical partition function for a one-dimensional gas of particles subject to fermionic-type exclusion rules. This  interpretation leads  to a positive multiple-sum 
expression of the characters \cite{OleJS}.

A natural question is the following: is there a natural operator formulation underlying this fermi-gas representation?


As a first guiding observation, let us recall (cf. \cite{BMlmp,OleJS, Kyoto,FWa}) that the  $\M(k+1,k+2)$ model is dual to  the $\z_k$ parafermionic theory \cite{ZF}. The duality is defined at the level of the weight function. As said before, the vertices that contribute to the weight of a $\M(k+1,k+2)$ path are those which do not correspond to local extrema and their contribution is 
$x/2$. Paths in the dual model are defined as in the original model but they are weighted in a dual way: only the extrema do contribute and they contribute to the value $x/2$. 
Each path is in correspondence with a state  of the finitized parafermionic module \cite{Path}. Since a character is a $q$-weighted sum over paths,  the duality between the weight function is lifted, at the level of characters, to a $q\lrw q^{-1}$ transformation (up to easily fixed $L$-dependent powers of $q$).

Now in the parafermionic case, the path representation is very close to the quasi-particle description of the conformal model \cite{LP,JMqp}. As demonstrated in \cite{Path}, each peak can be identified with the  mode  of a parafermionic operator whose parafermionic charge is  twice the  charge of the peak (to be defined below) and its mode index  is minus the $x$ position. 

A path is a sequence of peaks and, as we just indicated, in the parafermionic context this sequence can be interpreted as a string  of parafermionic  modes. 
The path itself is manifestly independent upon  the 
way it is weighted. It is thus tempting to guess that, loosely speaking,  the operator formulation underlying the path representation of   the minimal models could similarly be described by parafermionic-type operators but which would then, in the light of the  $q\lrw q^{-1}$ duality, acts in some non-local way.

This sets the stage for the present investigation. We search for an operator description of the $\M(k+1,k+2)$ paths and expect this to be, somehow, non-local.

\subsection{From paths to an operator basis}

Concretely, we look for a description of an irreducible $\M(k+1,k+2)$  module (specified by some boundary conditions) in terms of a string of operators acting  on the path with lowest energy (or weight)  within this module. We next lift the set of constraints fixed by the path on allowed sequences of such operators to an independent operator basis describing irreducible modules.

It turns out not to be difficult to work  out an operator interpretation of the $\M(k+1,k+2)$ paths. It first consists in interpreting each weight-contributing vertex as the insertion point of an operator of energy 
equal to the weight of the vertex; if the vertex has coordinates $(x,y)$, this weight is  $x/2$.  Roughly, one type of  operator acts at a peak position and  creates a NE edge linking $x$ and $x+1$. Let us call this operator $b_{x}$. By construction, two operators cannot act at the same point since after one action, a vertex that corresponds to a maximum is transformed into the starting point of a straight-up segment. This makes these operator of fermionic type ($b_{x}^2=0$). By creating a NE edge, $b_{x}$  lifts  the height of the whole tail of the path, from the point $x+1$ to its right end, by one unit. This is definitely a non-local action. It also clearly modifies the path  boundary condition at the right extremity.
The precise 
definition of this action is presented in section 2. 

Because each action  of a $b$ operator lifts the path tail, a specific  module cannot be described solely in terms of such $b$-type operators acting in all possible ways on the ground-state configuration of the module, since all paths pertaining to a given module must have the same end condition.  One also needs to have an operator that decreases the height of the path tail by creating a SE edge.
Let us call such an operator  $b^*$. 
 We thus see that the states in a given module (i.e., with fixed values of $(y_0,y_L)$) must be described by sequences containing equal numbers of $b$ and $b^*$ operators acting on the module's ground state. 
 Moreover, since the height of a path can never be larger than $k$, 
 there cannot be more that $k-1$ adjacent $b$ or $b^*$ operators
The full set of relations is presented in Section 2.


In the present article, we provide an application of this operator formalism, by re-deriving the finitized character of the unitary minimal models from its defining basis. This derivation  turns out to be somewhat more economical that the original one using the path description \cite{OleJS} -- although it is fair to stress that in its essence, it is not intrinsically different. But this analysis really serves as a verification of the correctness of the induced operator basis (i.e., the completeness of the set of restrictions). This makes this technique available for the investigation of more complicated problems (cf. the concluding remarks).

The operator construction presented here for the $\M(k+1,k+2)$ models can be  extended rather directly to the finite 
 $\M(k+1,2k+3)$ models. This special class of non-unitary models has been shown in \cite{JSTAT} to have a path description very similar to that of the  $\M(k+1,k+2)$ models except that the lattice is half-integer and peaks  are forced to be at integer $(x,y)$ positions. But the crucial point is that the weight function is defined exactly as for the unitary models: all vertices but the  extrema contribute to $x/2$. The operator basis is thus very similar to that for the unitary models. It is briefly described in Section 3. Note that the corresponding dual models are the graded $\z_k$ parafermions \cite{CRS,JM}.

Finally, we should point out  that a somewhat analogous operator construction  for the unitary minimal  models  has been considered before, in \cite{FP}. However, the main difference is  that these authors have considered uncharged fermionic operators, which makes the basic relations somewhat different (in particular the partial charge condition below in (\ref{cc}) is more elaborated here).
In addition, we claim to have an operator basis, hence no unphysical sequences get generated that would necessitate the introduction of  projectors as in \cite{FP}.

\section{Paths for $\M(k+1,k+2)$ models and their operator interpretation}

\subsection{Paths, charge and the fermionic character}
Let us start by recalling the definition of the  paths representing the $\M(k+1,k+2)$ models.
These are defined on  an integer lattice in the first quadrant  of the $(x,y)$ plane and within the rectangle $0\leq y \leq k$ and $0\leq x \leq L$. 
An edge from $x$ to $x+1$ is either NE or SE.
The {\it weight} $w$  of a  path is
\begin{equation}\label{weig}
w= \sum_{x=1}^{L-1} w(x)\qquad \text{where} \qquad w(x)= \frac{x}4 \big|\, y_{x+1}-y_{x-1}\, \big|.
\end{equation} 
In other words, $w(x)=0$ if $x$ corresponds to the horizontal coordinate of an extremum of the path and it is equal  to $x/2$ otherwise.

For a unique identification between the set of  paths with specific boundary conditions  and the states of a given module, we also need to specify the last edge (SE or NE) \cite{Mel, OleJS}. Here it will be  understood that all paths terminate with a SE edge. 
The generating function of paths with boundary conditions $(y_0,y_L)$ yields the finitized character $\chi_{r,s}^{(L)}(q)$ with 
\begin{equation}\label{rs}
r=y_L+1 \qquad \text{and} \qquad s=y_0+1.
\end{equation}  The condition $0\leq y_L\leq k-1$, which is forced by the SE edge termination, implies the right bound for $r$.\footnote{If we insist instead by terminating the path with a NE edge, the proper identification of the first label is $r=y_L$.} 

For simplicity, we mainly confine ourself to paths starting and terminating at height 0, which describe the states in the finitized vacuum module. Indeed, our aim is essentially  to dress the known fermionic formula with a new interpretation and for that it is best to avoid unnecessary complications induced by boundary effects. Therefore, in the rest of this section (except in the last subsection where the other modules are briefly considered), a path means a path with the particular condition $(y_0,y_L)=(0,0)$.

A path is thus a sequence of peaks
of height between 1 and $k$. The height by itself is not a good intrinsic characterization of the peaks. The proper concept is that of relative height \cite{BreL} or charge {\cite{OleJS}. We will use the latter qualitative. The definition is as follow.
 The charge of a peak with  coordinates $(x,y)$ is the largest number $c$ such that we can find two vertices $(x',y-c)$ and $(x'',y-c)$ on the  path  with $x'<x<x''$ and such that between these two vertices there are no peak of height larger than $y$ and every peak of height equal to $y$ is located at its right \cite{BP}. 
 
For an isolated peak described by a triangle of height $\ell$ starting and ending on the $x$ axis,  the charge is equal to the height. This is not as simple within a complex \cite{BreL,OleJS}, that is, within portions of the paths delimited by two points on the horizontal axis and containing more than one peak. In particular, if there are two peaks at the same height, the above definition indicates that 
it is the left-most peak which is attributed the largest charge, the one given by the height. In general thus, the charge of a peak is $\leq $ its height.
Fig.  1 illustrates the definition of the charge.
 

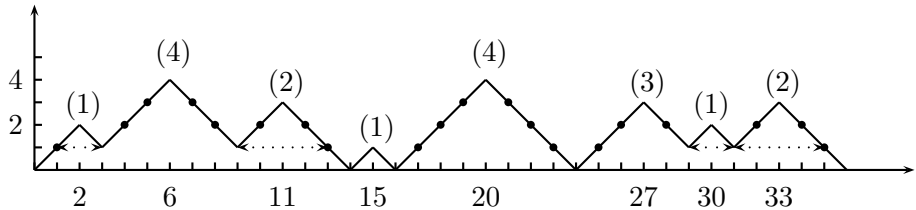
\begin{figure}[ht]
\caption{{\footnotesize {An example of a path valid for $k \geq 4$ starting at $y_0=0$ and ending at $y_{36}=0$.  There are four complexes in the path, delimitated by the $x$ positions 0, 14, 16, 24 and 36.  The charge of each  peak
is indicated in parenthesis. A dotted line indicates the line from which the height must be measured to give the charge. The charge content is thus $n_1=3,\, n_2=2,\, n_3=1$ and $n_4=2$. The total charge is   $\sum_i in_i= 18$, half the length of the path. The weight of this path is half the sum of the $x$-coordinate of the dotted vertices ($w=180)$. }}}\label{fig1}
\begin{center}
\begin{pspicture}(0,0)(12.5,3)
\psline{->}(0.3,0.3)(0.3,2.5) \psline{->}(0.3,0.3)(12.0,0.3)
\psset{linestyle=dotted} \psline{<->}(0.6,0.6)(1.2,0.6)
 \psline{<->}(9.6,0.6)(10.8,0.6)
\psline{<->}(3.0,0.6)(4.2,0.6) \psline{<->}(9.0,0.6)(9.6,0.6)
\psset{linestyle=solid}
\psline{-}(0.3,0.3)(0.3,0.4) \psline{-}(0.6,0.3)(0.6,0.4)
\psline{-}(0.9,0.3)(0.9,0.4) \psline{-}(1.2,0.3)(1.2,0.4)
\psline{-}(1.5,0.3)(1.5,0.4) \psline{-}(1.8,0.3)(1.8,0.4)
\psline{-}(2.1,0.3)(2.1,0.4) \psline{-}(2.4,0.3)(2.4,0.4)
\psline{-}(2.7,0.3)(2.7,0.4) \psline{-}(3.0,0.3)(3.0,0.4)
\psline{-}(3.3,0.3)(3.3,0.4) \psline{-}(3.6,0.3)(3.6,0.4)
\psline{-}(3.9,0.3)(3.9,0.4) \psline{-}(4.2,0.3)(4.2,0.4)
\psline{-}(4.5,0.3)(4.5,0.4) \psline{-}(4.8,0.3)(4.8,0.4)
\psline{-}(5.1,0.3)(5.1,0.4) \psline{-}(5.4,0.3)(5.4,0.4)
\psline{-}(5.7,0.3)(5.7,0.4) \psline{-}(6.0,0.3)(6.0,0.4)
\psline{-}(6.3,0.3)(6.3,0.4) \psline{-}(6.6,0.3)(6.6,0.4)
\psline{-}(6.9,0.3)(6.9,0.4) \psline{-}(7.2,0.3)(7.2,0.4)
\psline{-}(7.5,0.3)(7.5,0.4) \psline{-}(7.8,0.3)(7.8,0.4)
\psline{-}(8.1,0.3)(8.1,0.4) \psline{-}(8.4,0.3)(8.4,0.4)
\psline{-}(8.7,0.3)(8.7,0.4) \psline{-}(9.0,0.3)(9.0,0.4)
\psline{-}(9.3,0.3)(9.3,0.4) \psline{-}(9.6,0.3)(9.6,0.4)
\psline{-}(9.9,0.3)(9.9,0.4) \psline{-}(10.2,0.3)(10.2,0.4)
\psline{-}(10.5,0.3)(10.5,0.4) \psline{-}(10.8,0.3)(10.8,0.4)

\rput(0.9,-0.05){{\small $2$}} \rput(2.1,-0.05){{\small $6$}}
\rput(3.6,-0.05){{\small $11$}} \rput(4.8,-0.05){{\small
$15$}}\rput(6.3,-0.05){{\small $20$}} \rput(8.4,-0.05){{\small $27$}}
\rput(9.3,-0.05){{\small $30$}}\rput(10.2,-0.05){{\small $33$}}
 \psline{-}(0.3,0.6)(0.4,0.6)
\psline{-}(0.3,0.9)(0.4,0.9) \psline{-}(0.3,1.2)(0.4,1.2)
\psline{-}(0.3,1.5)(0.4,1.5) \psline{-}(0.3,1.8)(0.4,1.8)

\rput(0.05,0.9){{\small $2$}} \rput(0.05,1.5){{\small $4$}}
\psline{-}(0.3,0.3)(0.6,0.6) \psline{-}(0.6,0.6)(0.9,0.9)

\psline{-}(0.9,0.9)(1.2,0.6)

\psline{-}(1.2,0.6)(1.5,0.9) \psline{-}(1.5,0.9)(1.8,1.2)
\psline{-}(1.8,1.2)(2.1,1.5)

\psline{-}(2.1,1.5)(2.4,1.2) \psline{-}(2.4,1.2)(2.7,0.9)

\psline{-}(2.7,0.9)(3.0,0.6) \psline{-}(3.0,0.6)(3.3,0.9)

\psline{-}(3.3,0.9)(3.6,1.2) \psline{-}(3.6,1.2)(3.9,0.9)
\psline{-}(3.9,0.9)(4.2,0.6)

\psline{-}(4.2,0.6)(4.5,0.3)

\psline{-}(4.5,0.3)(4.8,0.6)

\psline{-}(4.8,0.6)(5.1,0.3) \psline{-}(5.1,0.3)(5.4,0.6)
\psline{-}(5.4,0.6)(5.7,0.9)

\psline{-}(5.7,0.9)(6.0,1.2)\psline{-}(6.0,1.2)(6.3,1.5)
\psline{-}(6.3,1.5)(6.6,1.2) \psline{-}(6.6,1.2)(6.9,0.9)

\psline{-}(6.9,0.9)(7.2,0.6) \psline{-}(7.2,0.6)(7.5,0.3)

\psline{-}(7.5,0.3)(7.8,0.6) \psline{-}(7.8,0.6)(8.1,0.9)

\psline{-}(8.1,0.9)(8.4,1.2) \psline{-}(8.4,1.2)(8.7,0.9)

\psline{-}(8.7,0.9)(9.0,0.6)

\psline{-}(9.0,0.6)(9.3,0.9) \psline{-}(9.3,0.9)(9.6,0.6)
\psline{-}(9.6,0.6)(9.9,0.9)

\psline{-}(9.9,0.9)(10.2,1.2) \psline{-}(10.2,1.2)(10.5,0.9)
\psline{-}(10.5,0.9)(10.8,0.6) \psline{-}(10.8,0.6)(11.1,0.3)
\psset{dotsize=3pt}
\psdots(0.6,0.6)(1.5,0.9)(1.8,1.2)(2.4,1.2)(2.7,0.9)(3.3,0.9)(3.9,0.9)
(4.2,0.6)(5.4,0.6)(5.7,0.9)(6.0,1.2)(6.6,1.2)(6.9,0.9)(7.2,0.6)(7.8,0.6)
(8.1,0.9)(8.7,0.9)(9.9,0.9)(10.5,0.9)(10.8,0.6)

\rput(0.9,1.15){{\small { ${(1)}$}}}
\rput(2.1,1.85){{\small { $(4)$}}}
\rput(3.6,1.45){{\small {  $(2)$}}}
\rput(4.8,0.85){{\small { $(1)$}}}
\rput(6.3,1.85){{\small { $(4)$}}}
\rput(8.4,1.45){{\small { $(3)$}}}
\rput(9.3,1.15){{\small {$(1)$}}}
\rput(10.2,1.45){{\small { $(2)$}}}

\end{pspicture}
\end{center}
\end{figure}


A path is fully  determined  by the specification of its peak positions and their charges. In particular, the (vacuum) ground-state path for all unitary minimal models  is described by a sequence of $L/2$ peaks of charge 1 -- see Fig. 2a.

Denote by $n_j$ the number of peaks of charge $j$. 
The length of a path is twice the sum of the charge of all its peaks
\begin{equation} 
L=\sum_{j=1}^k 2j n_j\ .\end{equation} 
Note that $L$ is always even.

The values of $n_j$  are the very basic numbers that enters in the fermionic expression of the characters. This expression, in its finitized form, reads \cite{OleJS}
\begin{equation} \label{chi11}
\chi_{1,1,}^{(L)}(q) =\sum_{n_1,\ldots, n_k\geq 0} q^{nBn}\;\prod_{j=1}^{k-1}
\begin{bmatrix}
n_j+m_j\\ n_j\end{bmatrix} ,
\end{equation} 
with 
\begin{equation}\label{bij}
 nBn= \sum_{i,j=2}^{k-1}n_i B_{ij} n_j, \quad \text{where}\quad B_{ij}=B_{ji}, \qquad B_{ij}=(i-1)j\quad \text{if} \quad i\leq j,
\end{equation} 
and
\begin{equation} m_j=2n_{j+1}+4n_{j+2}+\cdots +2(k-j)n_k.
\end{equation}

\subsection{Operators acting on paths: setting the basis} 

The first  objective is to describe any path by the successive action  of operators acting on the ground-state configuration. We will thus interpret every weight-contributing vertex as resulting from the insertion of an operator. Next, we will determine, from the path characteristics, those conditions on sequences of these operators  that capture  the elements of an operator basis.

As already indicated in the introduction, this construction requires two types of operators, $b$ and $b^*$, that  create a NE and a SE edge respectively (the precise action is defined below). To these  operators, we associate the respective  charge $+1$ and $-1$.  Moreover, the mode of an operator corresponds to  its horizontal insertion point. For instance $b_{5}$ indicates that the operator $b$ acts at the vertex at horizontal position 5. The weight of an operator is half the value of its mode. The weight of a sequence of operators is thus the sum of the mode indices within the sequence divided by two. 

\let\a\alpha

\let\lrw\leftrightarrow
 In order to define the action of these operators in a neat way, we observe that a path is determined by the sequence of its edges. With NE $\lrw 1$ and SE $\lrw-1$,
it can thus be regarded as a succession of $\pm1$:
  \begin{equation} \text {Path}: \quad {\bf \a}\equiv  (\a_1,\a_2,\cdots , \a_L)\qquad \text{with}\quad \a_i\in\{1,-1\} \quad \text{and}\quad \a_1=-\a_L=1.
\end{equation}
The edge $\a_i$ links the vertices at $x$-position
 $i-1$ and $i$.
The restriction on the height takes the form:
 \begin{equation}\label{rest} 
 0\leq h_i \leq k\qquad \text{where }\qquad  h_i\equiv\sum_{j=1}^i\a_j.
\end{equation}
 The operator $b_i$ acts on the vertex $i$, or equivalently, in-between the edges $\a_i$ and $\a_{i+1}$. Its action is defined as:
 \begin{multline}
\qquad b_i  (\a_1,\cdots, \a_i,\a_{i+1},\cdots , \a_L) = \\
(\a_1,\cdots, \a_i,1,\a_{i+1},\cdots , \a_{L-1})\, \delta_{\a_i,1}\, \delta_{\a_{i+1},-1}\, \chi(h_i\leq k-1)
\end{multline} 
where $\chi(a)$ =1 if $a$ is true and zero otherwise.
 Let us rephrase this definition in words. 
  Being designed to create a NE edge (which is the 1 in $i+1$-th position of the word), this operator $b$ must act on a peak, i.e., in-between a pair of edges of type $(1,-1)$. This accounts for the two delta function. By creating  a NE edge, this action of $b$ manifestly affects the part of the path at the right of its insertion point. 
Its effect is as follows: it translates the remaining part of the path by one unit upward and one unit toward the right. Moreover, in order to be defined for $L$ fixed, it removes the last edge. Finally,  the restriction defined by the truth function takes care of the restriction (\ref{rest}). 
The action of the operator $b_{1}$ on the path representing the ground state is described in Fig. 2b.

A number of immediate consequences follow from this definition.   On any path $\a$ such that $b_i\, \a= \a'=(\a'_1\cdots , \a_L')\not=0$, then
\begin{align}\label{bb}
(1)\qquad  & b_i^2\a=0,&\\ \nonumber
(2)\qquad  &b_{i+2}b_i\, \a =0&\\  \nonumber
(3)\qquad & b_{i+k}\cdots b_{i+1}b_i\a =0.
\end{align}
Indeed, if $ \a'\not=0$, then $\a_{i+1}'=1$ so that $b_i\, \a'\propto \delta_{\a'_{i+1},-1}=0$.
Similarly,  $\a_{i+2}'=-1$ so that $b_{i+2}\, \a'\propto \delta_{\a'_{i+2},1}=0$. Finally, 
   $b_{i+k}\cdots b_{i+1}\a' \propto \chi(h_i\leq 0)=0$ since $ \a'\not=0$ implies that $h_i\geq 1 $.
These three relations capture the three constraints in the definition of the action of $b_i$ specified by the two delta functions and the truth function.

\begin{figure}[ht]
\caption{{\footnotesize {The ground state pertaining to the vacuum module of all the unitary minimal models is displayed in (a) for $L=12$.  The action of the operator  $b_{1}$ on the ground state is presented in (b). This action is clearly non-local, modifying the path form the insertion point, $x=1$, to the end, $x=12$. One edge NE  is created, linking $x=1$ to $x=2$, and the last SE edge of the original ground-state path is removed. The right boundary condition is clearly modified (the value of $y_{12}$ being now 2 instead of 0).  The action of the operator $ b^*_{3}$ on the path in (b) is given in (c). This action is again manifestly non-local: by creating a SE  edge from $x=3$ to $x=4$, it modifies the whole path from the right of $x=3$ and remove the final NE  edge. The path illustrated in  (c) is thus equivalent to the sequence  $b^*_{3}b_{1}$ acting on the ground state configuration. }}} \label{fig2}
\begin{center}
\begin{pspicture}(0,0)(7.0,5.0)
\psline{->}(0.5,0.5)(7.0,0.5) \psline{->}(0.5,2.0)(7.0,2.0)
\psline{->}(0.5,3.5)(7.0,3.5)

\psline{->}(0.5,0.5)(0.5,1.5)\psline{->}(0.5,2.0)(0.5,3.0)
\psline{->}(0.5,3.5)(0.5,4.5)

\psset{linestyle=solid}

\psline{-}(0.5,0.5)(0.5,0.6) \psline{-}(1.0,0.5)(1.0,0.6)
\psline{-}(1.5,0.5)(1.5,0.6) \psline{-}(2.0,0.5)(2.0,0.6)
\psline{-}(2.5,0.5)(2.5,0.6) \psline{-}(3.0,0.5)(3.0,0.6)
\psline{-}(3.5,0.5)(3.5,0.6) \psline{-}(4.0,0.5)(4.0,0.6)
\psline{-}(4.5,0.5)(4.5,0.6) \psline{-}(5.0,0.5)(5.0,0.6)
\psline{-}(5.5,0.5)(5.5,0.6) \psline{-}(6.0,0.5)(6.0,0.6)
\rput(2.5,0.25){{\small $4$}}\rput(4.5,0.25){{\small
$8$}}\rput(1.5,0.25){{\small $2$}}\rput(3.5,0.25){{\small $6$}}
\rput(5.5,0.25){{\small $10$}}\rput(6.5,0.25){{\small $12$}}

\psline{-}(0.5,2.0)(0.5,2.1) \psline{-}(1.0,2.0)(1.0,2.1)
\psline{-}(1.5,2.0)(1.5,2.1) \psline{-}(2.0,2.0)(2.0,2.1)
\psline{-}(2.5,2.0)(2.5,2.1) \psline{-}(3.0,2.0)(3.0,2.1)
\psline{-}(3.5,2.0)(3.5,2.1) \psline{-}(4.0,2.0)(4.0,2.1)
\psline{-}(4.5,2.0)(4.5,2.1) \psline{-}(5.0,2.0)(5.0,2.1)
\psline{-}(5.5,2.0)(5.5,2.1) \psline{-}(6.0,2.0)(6.0,2.1)

\psline{-}(0.5,3.5)(0.5,3.6) \psline{-}(1.0,3.5)(1.0,3.6)
\psline{-}(1.5,3.5)(1.5,3.6) \psline{-}(2.0,3.5)(2.0,3.6)
\psline{-}(2.5,3.5)(2.5,3.6) \psline{-}(3.0,3.5)(3.0,3.6)
\psline{-}(3.5,3.5)(3.5,3.6) \psline{-}(4.0,3.5)(4.0,3.6)
\psline{-}(4.5,3.5)(4.5,3.6) \psline{-}(5.0,3.5)(5.0,3.6)
\psline{-}(5.5,3.5)(5.5,3.6) \psline{-}(6.0,3.5)(6.0,3.6)

\psline{-}(0.5,3.5)(1.0,4.0) \psline{-}(1.0,4.0)(1.5,3.5)
\psline{-}(1.5,3.5)(2.0,4.0) \psline{-}(2.0,4.0)(2.5,3.5)
\psline{-}(2.5,3.5)(3.0,4.0) \psline{-}(3.0,4.0)(3.5,3.5)
\psline{-}(3.5,3.5)(4.0,4.0) \psline{-}(4.0,4.0)(4.5,3.5)
\psline{-}(4.5,3.5)(5.0,4.0) \psline{-}(5.0,4.0)(5.5,3.5)
\psline{-}(5.5,3.5)(6.0,4.0) \psline{-}(6.0,4.0)(6.5,3.5)
\rput(-0.5,4.0){{\small a)}}

\psline{-}(0.5,2.0)(1.0,2.5) \psline{-}(1.0,2.5)(1.5,3.0)
\psline{-}(1.5,3.0)(2.0,2.5) \psline{-}(2.0,2.5)(2.5,3.0)
\psline{-}(2.5,3.0)(3.0,2.5) \psline{-}(3.0,2.5)(3.5,3.0)
\psline{-}(3.5,3.0)(4.0,2.5) \psline{-}(4.0,2.5)(4.5,3.0)
\psline{-}(4.5,3.0)(5.0,2.5) \psline{-}(5.0,2.5)(5.5,3.0)
\psline{-}(5.5,3.0)(6.0,2.5)\psline{-}(6.0,2.5)(6.5,3.0)
\rput(-0.5,2.5){{\small b)}}

\psline{-}(0.5,0.5)(1.0,1.0) \psline{-}(1.0,1.0)(1.5,1.5)
\psline{-}(1.5,1.5)(2.0,1.0) \psline{-}(2.0,1.0)(2.5,0.5)
\psline{-}(2.5,0.5)(3.0,1.0) \psline{-}(3.0,1.0)(3.5,0.5)
\psline{-}(3.5,0.5)(4.0,1.0) \psline{-}(4.0,1.0)(4.5,0.5)
\psline{-}(4.5,0.5)(5.0,1.0) \psline{-}(5.0,1.0)(5.5,0.5)
\psline{-}(5.5,0.5)(6.0,1.0)\psline{-}(6.0,1.0)(6.5,0.5)
\rput(-0.5,1.0){{\small c)}}

\psset{dotsize=3pt}
\psdots(1.0,2.5)(1.0,1.0)(2.0,1.0)

\end{pspicture}
\end{center}
\end{figure}
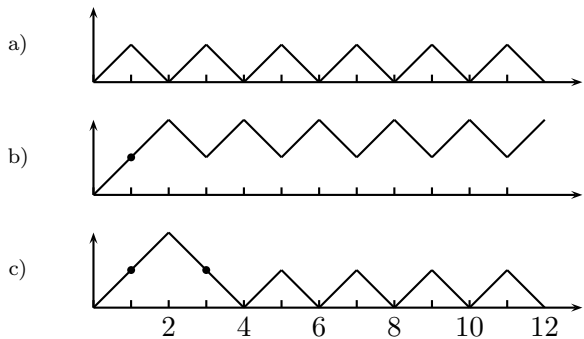


After the action of $\ell$ operators of $b$-type, the boundary condition $y_L=0$ is changed to $y_L= 2\lceil \ell/2\rceil$. To preserve the original boundary condition, we need to act with a sequence of operators $b^*$ such that each $b^*$ creates a SE edge.  Actually, for the end of the path  to reach the horizontal axis, there must be as many $b^*$ as $b$.

The precise definition of the action of the operator $b^*_i$ is:
 \begin{multline}
\qquad 
b^*_i \, (\a_1,\cdots, \a_i,\a_{i+1},\cdots , \a_L) = (\a_1,\cdots, \a_i,-1,\a_{i+1},\cdots , \a_{L-1})\, \delta_{\a_i,-1}\, \delta_{\a_{i+1},1}\, \chi(h_i\geq 1).
\end{multline} 
This implies  
\begin{align}\label{bsbs}
(1)\qquad  & {b_i^*}^2\a=0,&\\ \nonumber
(2)\qquad  &b^*_{i+2}b^*_i\, \a =0&\\  \nonumber
(3)\qquad & b^*_{i+k}\cdots b^*_{i+1}b^*_i\a =0.
\end{align}
It also follows directly from the previous two definitions that
\begin{equation}\label{bbs}
b_{i} b^*_{i}\a=b^*_{i} b_{i}\a=0
\quad \text{and}\quad b_{i+1} b^*_{i}\a=b^*_{i+1} b_{i}\a=0.
\end{equation}

 
Let us now use these operators to set up a basis of states in the vacuum module which, in its final formulation, would be path-independent. For this we impose as constraints the three equalities in eqs (\ref{bb}) and (\ref{bsbs}). Note  that on the ground-state path of Fig \ref{fig2}a, the action of adjacent identical operators $b_jb_i$ or $b_j^*b_i^*$ must satisfy $j-i=1+2n$ for $n$ a positive integer (since the action is either successive, $j-i=1$, or further separated by a sequence of  peak of charge 1, hence by an even number of $x$ steps). For different operators, $b_jb^*_i$ or $b_j^*b_i$, the condition is rather $j-i=2+2n$. 
 
To these conditions, we need to impose a constraint that reflects the fact that the height of a vertex has to be everywhere $\leq k$ and always $\geq 0$. This has been partly taken into account by the condition (3) of eqs (\ref{bb}) and (\ref{bsbs}). However,  these conditions are not restrictive enough as they allow, for example, a sequence of $\ell$ $(\leq k-1)$ $b$, followed by a $b^*$ and then a sequence of $\ell'$ $b$, with $\ell+\ell'\geq k+2$, which would lift a vertex to a height larger than $k$. Such possibilities need to be excluded. This forces the charge computed from the beginning to any point in the sequence of operators (the sequence being read from left to right)
to lie within the range  $[0,\,k-1]$. This can be made more precise as follows. Consider any sequence of modes of the form $c_{i_m}c_{i_{m-1}} \cdots c_{i_1} $, with $c$ being either $b$ or $b^*$,  and denote  the charge by $q$ (i.e.,  $q(b)=-q(b^*)=1$). Introduce the partial charge  $q_\ell$ as
\begin{equation}\label{ql}  q_\ell= \sum_{n=1}^\ell q(c_{i_n}).
\end{equation}
Then, to ensure that $0\leq y\leq k$ everywhere, one requires all partial charges to satisfy 
 \begin{equation}\label{cc}
0\leq q_\ell\leq k-1 \qquad \quad (1\leq \ell\leq m). 
\end{equation}
We stress that (\ref{cc})  embodies a non-locality aspect: it can be viewed as a constraint on the type of operator to-be-inserted  which involves not only the preceding operator, but, in principle, all those which have  been previously introduced.\footnote{The constraint on successive allowed modes (cf. (3) in eqs (\ref{bb}) and (\ref{bsbs}))  can be viewed as a sort of weak form of the exclusion principle that governs the quasi-particle  parafermionic basis (as it appears in the form of the `difference 2 condition at distance $k-1$' \cite{LP,JMqp}). But the above  specific bound on the charge of strings of $b$ and $b^*$ modes  is a novelty with regard to constraints on the charge in parafermionic theory  since the parafermionic charge -- when used to organize the modules --, is in fact defined modulo $2k$, so that no states are really forbidden by the requirement that the  relative charge $r$ should be  reshuffled  within the bounds $-k< r\leq k$. Viewed  differently, the charge of a string of operators preceding the application of a given $b$ or $b^*$ acts as a sort of exclusion principle while in the parafermionic context, this is not so; it is true that the charge up to the point of application of a given parafermionic mode does play a role but only as a shift of the mode index. }

The set of conditions identified completely define the operator basis proper to the vacuum module. These conditions can be summarized as follows:
\begin{align}\label{basis}  c_{i_m}\cdots c_{i_2}  c_{i_1} :&\quad \text{$c$ is either $b$ or $b^*$, with $q(b)=-q(b^*)=1$,}\nonumber 
\\
& \quad i_{s+1}-i_s = 2n+ 1+\frac12(1-q(c_{i_{s+1}})\, q(c_{i_s})), \nonumber\\ &\quad\text{with $i_1= 1+2n'$ and $i_m\leq L-1$},\nonumber\\ &  \quad 0\leq q_\ell\leq k-1,\quad \text{with $q_1=1$ and $q_m=0$},
\end{align}
where $n$ and $n'$ are non-negative integer. The finitized (vacuum) character is the generating function of all the sequences of operators subject to (\ref{basis}).
Its construction is considered in the next two subsections. At the end, the correctness of the resulting expression testifies that of the basis (\ref{basis}).

\subsection{Combinatorics of the operator basis: the minimal-weight configuration  and its displacements}

Our aim is now to derive the generating function of all the strings of operators satisfying the conditions (\ref{basis}).
%
The first step is the identification of a convenient choice of summation variables. For paths, it is the charge content of the peaks, that is, the different $n_j$,  which provides the natural summation variables. The analogous variables for strings of operators are readily identified.

 Let us call a sequence of $2\ell$ operators ordered (with decreasing value of their mode, i.e., position action) as $b^*\cdots b^*b\cdots b$ a $\ell$-block.
 Denote by $p_\ell$ the number of $\ell$-blocks. By analogy with paths, the operator content over which the summation will be performed is thus the block content, that is, the values of $p_1,\cdots, p_{k-1}$.

The problem can thus be formulated more precisely as follows: (1)- enumerate all possible sequences with a specified block content, and (2)- sum over all block contents compatible with the finitization.

 It is rather immediate to see that 
among all sequences of operators with a specific block content, the pattern with minimal weight is the one with all blocks unmixed\footnote{The effect of mixing the blocks is considered in the next section and shown there to increase the weight with respect to the (unmixed) configuration considered here.}, as closely packed as possible, and ordered in decreasing values of $\ell$ (from right to left).
Rephrased more explicitly, this state is described by sets of operators with modes as closely packed as allowed by the hard-core repulsion (\ref{basis}) and ordered as follows, from  right to left to: the $p_{k-1}$ $(k-1)$-blocks, followed by the $p_{k-2}$ $(k-2)$-blocks, down to the the $p_{1}$ blocks of type 1.

The rationale for identifying this particular operator sequence with the lowest-weight state is that the larger the value of $\ell$ (the block type), the greater is the number of $b$ operators contributing to the weight so that it is energetically favorable to have them in lowest available modes. Note that even in a closely-packed configuration, the ordering of the blocks matters significantly with regard to the total weight since there is a gap of 1 in-between every pair of adjacent $bb^*$ and $b^*b$.

 Let us now determine the weight of the minimal-weight configuration. Consider the closely-packed $\ell$-block of the form $b^*_{i_0+2\ell+1}\cdots b^*_{i_0+\ell+2}\, b_{i_0+\ell}\cdots  \,b_{i_0+1}$.  Its total weight is 
\begin{equation}
\frac12\sum_{j=1}^{2\ell+1}(i_0+j) - \frac12(i_0+\ell+1) =\ell(i_0+\ell+1),
\end{equation}
where the subtraction keeps track of the gap of 1 between $b^*_{i_0+\ell+2}$ and $b_{i_0+\ell}$. For a sequence of $p_\ell$ such blocks, one has:
\begin{equation}
\frac12\sum_{j=1}^{2(\ell+1)p_\ell}(i_0+j) - \frac12\sum_{j=1}^{2p_\ell}(i_0+j(\ell+1)) =\ell\, p_\ell(i_0+(\ell+1)p_\ell).
\end{equation}
This clearly indicates that the largest the value of $\ell$, the more energetic it is to translate $\ell$-blocks. The value of $i_0$ is the sum of the diameter of all higher blocks in their closely-packed form:
\begin{equation}
i_0 = 2\sum_{j=\ell+1}^{k-1}(j+1)p_j.
\end{equation}
The minimal-weight configuration with specified block content $(p_1,\cdots, p_{k-1})$ is thus  found to be
\begin{equation}
\sum_{i,j=1}^{k-1} p_i [\text{min}(i,j)(\text{max}(i,j)+1)]= \sum_{i,j=1}^{k-1} p_i B_{i+1, j+1} p_j ,
\end{equation}Ê
where $B_{ij}$ is defined in (\ref{bij}).

Consider now all possible translations of this configuration. By this we mean all possible displacements of the various operators which preserve their ordering. The displacement of each operator is measured from its position in the minimal-weight configuration.
Consider first the  leftmost operator, which is of type $b^*$. Its mode is $M-1$, with $M$ defined by  
 \begin{equation}\label{defm}
 M= 2\sum_{j=1}^{k-1}(j+1)\, p_j.
\end{equation} 
$M$ is  the sum of the diameter  of all the blocks.
The operator $b^*_{M-1}$ can be translated, in units of 2,  up to $L-1$. Its maximal displacement is thus $L-M$ and the weight increase is  $(L-M)/2$. 

The penultimate operator can similarly be translated by an even integer which is not larger than the displacement of the last operator. More generally, the different displacements
of the various operators, taken from  left to right, are in correspondence with a  sequence of numbers, which, by construction (since the ordering of the operator must remain unaffected) are non-increasing.   Denote these different numbers by $(2\mu_1,\cdots , 2\mu_P)$, where $2\mu_i$ is the displacements of the $i$-th operator counted from the left, so that $\mu_{i}\geq \mu_{i+1}$ and  $\mu_1\leq (L-M)/2$.  $P$ is 
the total number of operators:
\begin{equation}\label{defP}
 P=2\sum_{j=1}^{k-1}j p_j.
 \end{equation}
Note that some of these $\mu_j$ (necessarily those at the end of the sequence) are allowed to vanish.

We now argue that the weight increase resulting from such displacements is
\begin{equation}
n=\mu_1+\cdots +\mu_P.
 \end{equation}
Consider first the weight increase caused by the displacement of the last
operator. A shift by $2\mu_1$, i.e.
\begin{equation} b^*_{M-1} \cdots \quad \rw \quad b^*_{M-1+2\mu_1} \cdots,
\end{equation}
 clearly increases the weight by $\mu_1$. Consider now a shift of the penultimate operator. Since it starts from a closely-packed configuration and that the ordering needs to be preserved, a shift of $2\mu_2$ of this operator must be accompanied by a shift of the same amount of the last operator by $2\mu_2$. The weight increase is $2\mu_2$. If the last operator is then further shifted by $2(\mu_1-\mu_2)$, augmenting thereby the weight by $\mu_1-\mu_2$, the total weight increase is $\mu_1+\mu_2$. By iterating this argument, we arrive at the claimed result.

Counting the possible displacements, 
while keeping track of the corresponding  weight increase,
amounts to evaluating the following multiple summation:
\begin{equation} \sum_{\mu_1=0}^{(L-M)/2}\sum_{\mu_2=0}^{\mu_1}\cdots \sum_{\mu_P=0}^{\mu_{P-1} }q^{\mu_1+\cdots + \mu_{P}}.
\end{equation} 
This is  equivalent to enumerate the  partitions $(\mu_1,\cdots,\mu_{P})$ of $n= \mu_1+\cdots + \mu_{P}$  into at most $P$  parts, each part being at most equal to $(L-M)/2$ and with each  partition being weighted by $q^n$. 
The generating function for the number of partitions $p(r,m,n)$ of $n$ into at most $m$ parts each $\leq r$ is 
 (cf.  \cite{Andr} Theorem 3.1):
\begin{equation} \sum_{n\geq 0} p(r,m,n)\, q^n 
= \begin{bmatrix}
r+m\\ m\end{bmatrix},
\end{equation} 
where the $q$-binomial coefficient is 
\begin{equation}
\begin{bmatrix}
a\\ b\end{bmatrix}=\frac{(q)_a}{(q)_{a-b}(q)_b},
\end{equation} and
with the $q$-factorial function $(q)_a$ being defined as
\begin{equation} (q)_a= (1-q)\cdots (1-q^a).
\end{equation} 
One thus concludes that:
\begin{equation}  \sum_{\mu_1=0}^{(L-M)/2}\sum_{\mu_2=0}^{\mu_1}\cdots \sum_{\mu_P=0}^{\mu_{P-1} }q^{\mu_1+\cdots+ \mu_{P}}=
\begin{bmatrix}
L/2-M/2+P\\ P\end{bmatrix}.
\end{equation} 
Let us define the integer $p_0$ as $L-M=2p_0 $ (so that $p_0$ is the maximal possible value of $\mu_1$). We can thus write
\begin{equation}\label{defL}
 L=2\sum_{j=0}^{k-1}(j+1)p_j.
\end{equation} 
Then the $q$-binomial reads
\begin{equation} \begin{bmatrix}
L/2-M/2+P\\ P\end{bmatrix} = \begin{bmatrix}
p_0+P\\ P\end{bmatrix}.
\end{equation} 


\subsection{Combinatorics of the operator basis: mixing the blocks}

Up to this point we have only considered those displacements that preserve the ordering of the operators as determined  by the minimal-weight configuration. The rationale for maintaining the ordering is the hard-core repulsion between the individual operators (cf. (\ref{basis})). However, this leaves open the possibility of moving an operator through a charge-less combination, that is, through a block. Equivalently phrased, we must keep track of the  possibility  of inserting small blocks within larger ones. 

The operation of mixing the blocks while maintaining the block content  raises the issue of determining the block content of a generic sequence of operators. This can be done simply, by identifying the blocks successively, from the smallest to the largest ones.
One thus  first identifies all  the 1-bocks, that is, all the pairs $b^*b$ within subsequences of the form $b^{i} b^*b b^{*j}$, with max $(i,j)>0$. Remove all these pairs; their number is the value of $p_1$ for the sequence under study. Proceed then similarly to the identification of the 2-blocks. Iterate the procedure until all $(k-2)$-blocks have been identified. The remaining operators form $(k-1)$-blocks.
For instance (assuming the modes to be increasing from right to left), for  the sequence
\begin{equation}
b^*b^*b^*b^*b^*bbbb^*bb^*b^*b^*b^* bbbb^*b^*bbb^*     bbbb,
\end{equation}
we have the following successive pairings (where the subscript indicates the content of the enclosed  block, hence the step at which the corresponding grouping is performed) 
\begin{equation}
b^*b^* (b^*b^*b^*bbb)_3(b^*b)_1 b^*b^*b^*b^* bbb (b^*b^*bb)_2 (b^*b)_1 
bbb
\end{equation}
The remaining operators form a 6-block. 
%
%
The complete content of this sequence  is $p_1=2,\, p_2=p_3=p_6=1$.

The first step in the analysis of mixing the blocks is to identify all possible (but distinct) ways of inserting a $j$-block within a $\ell$-block, for $j<\ell$. 
 It is understood that the block content (which at this point is still fixed) must be preserved by this procedure.
In this regard, what has to be prevented is the creation, from the mere mixing,  of blocks larger than $\ell$. This is most conveniently ensured by imposing a constraint on the partial charges. For every insertion, we need to verify that the partial charge satisfies (in this context) $0\leq q_n\leq \ell$, with $1\leq n\leq 2(j+\ell)$.

Before turning to the general case, it is convenient to treat a simple example, say $(j,\ell)=(1,2)$. One starts with $(b^*b)_1(b^*b^*bb)_2$ and move the  pair $b^*b$ within the 2-block. The possible cases, including the original configuration, are:
 \begin{equation}
(b^*b)b^*b^*bb,\quad  
 b^*(b^*b)b^*bb,\quad \quad b^*b^*(b^*b)bb,\quad b^*b^*b(b^*b)b,\quad
 b^*b^*bb (b^*b),
\end{equation}
(with the original displaced pair being identified by parentheses).
The third configuration has to be excluded since $q_3$ is 3, which is not within the range allowed here, namely $[0,2]$. 
(Equivalently, its decomposition in  blocks does not preserve the block content $p_1=p_2=1$; it rather corresponds to $p_1=p_2=0, \, p_3=1$.)
 Removing the parentheses, we see that the second and the fourth cases have the same structure. The corresponding pattern must be counted only once. We are left with 3 possibilities:
\begin{equation}
 b^*bb^*b^*bb,\quad   b^*b^*bb^*bb,\quad b^*b^*bbb^*b.
\quad
\end{equation}
Inserting the modes appropriate for the closely-packed version yields:
\begin{equation}\label{ex3}
  b_9^*b_7b_5^*b_4^*b_2b_1,\quad   b_9^*b_8^*b_6b_4^*b_2b_1,\quad b_9^*b_8^*b_6b_5b_3^*b_1
  .\end{equation}
The respective weights are $14,15$ and $16$. In other words, the distinct insertions of a 1-bock  $b^*b$ within a 2-block modify the weight by steps of 1 and there are three distinct configurations. With the weight evaluated with respect to the minimal-weight configuration, the three possibilities are taken care by the $q$-binomial
\begin{equation} \begin{bmatrix}
3\\ 1\end{bmatrix}= 1+q+q^2.
\end{equation} 

In order to see easily that the weight is increased by one for each successive displacements of the pair $b^*b$ within the 2-block, let us introduce the following pictorial devise.
Indicate by  $\bullet$ an occupied mode and by $\circ$ an empty one -- a hole -- (representing the gap in-between a $bb^*$ or a $b^*b$ pair). For the three configurations in (\ref{ex3}), we have the representations given in Fig. {\ref{figA}.

\begin{figure}[ht]
\caption{A pictorial representation of the  three configurations corresponding to the mixing of a 1-block into a 2-block. }
\vskip-0.5cm
\label{figA}
\begin{center}
\begin{pspicture}(0,0)(3.0,1.5)

{\psset{dotstyle=*,xunit=10pt,yunit=10pt}
\psdots(0,3)(2,3)(4,3)(5,3)(7,3)(8,3)
\psdots(0,2)(1,2)(3,2)(5,2)(7,2)(8,2)
\psdots(0,1)(1,1)(3,1)(4,1)(6,1)(8,1)
\psline(0.7,1.6)(1.3,1.6)
\psline(2.7,1.6)(3.3,1.6)
\psline(3.7,.6)(4.3,.6)
\psline(5.7,.6)(6.3,.6)
}
{\psset{dotstyle=o,xunit=10pt,yunit=10pt}
\psdots(1,3)(3,3)(6,3)
\psdots(2,2)(4,2)(6,2)
\psdots(2,1)(5,1)(7,1)
}

\end{pspicture}
\end{center}
\end{figure}
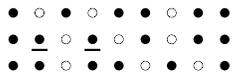
In this representation, a configuration is obtained from the previous one by the displacement toward the left of two dots separated by a hole. These displaced dots must be free to move together by one unit and their displacement should not generate a configuration containing three successive dots -- which would necessarily violate the charge constraint since no gaps between filled points means that they are all occupied by the same type of operators. The dots which have been moved are underlined. From this pictorial description, it is manifest that the weight is increased by one at each step.

As another  example, consider $p_1=2 $ and $p_2=1$. We have the 6 distinct  configurations displayed in Fig. \ref{figB} (where again the displaced dots  are underlined), 
together with their relative weight at their right. 
 From the second  configuration (of relative weight 1), there are two ways of displacing two dots by one unit and this leads to the two configurations of relative weight 2. These different possibilities are captured by the factor 
\begin{equation} \begin{bmatrix}
4\\ 2\end{bmatrix}= 1+q+2q^2+q^3+q^4.
\end{equation}
With $p_1$ 1-blocks and $p_2$ 2-blocks, the proper $q$-binomial factor takes the form
\begin{equation} \begin{bmatrix}
p_1+2p_2\\ p_1\end{bmatrix}.
\end{equation}

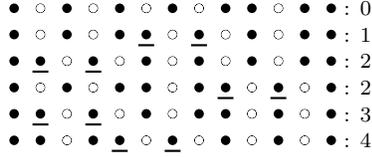
\begin{figure}[ht]
\caption{The configurations describing the mixing of a 1-block into two 2-blocks.}
\label{figB}
\begin{center}
\begin{pspicture}(0,0)(5.0,2.5)

{\psset{dotstyle=*,xunit=10pt,yunit=10pt}
\psdots(0,6)(2,6)(4,6)(6,6)(8,6)(9,6)(11,6)(12,6)
\psdots(0,5)(2,5)(4,5)(5,5)(7,5)(9,5)(11,5)(12,5)
\psdots(0,4)(1,4)(3,4)(5,4)(7,4)(9,4)(11,4)(12,4)
\psdots(0,3)(2,3)(4,3)(5,3)(7,3)(8,3)(10,3)(12,3)
\psdots(0,2)(1,2)(3,2)(5,2)(7,2)(8,2)(10,2)(12,2)
\psdots(0,1)(1,1)(3,1)(4,1)(6,1)(8,1)(10,1)(12,1)
\psline(4.7,4.6)(5.3,4.6)
\psline(6.7,4.6)(7.3,4.6)
\psline(0.7,3.6)(1.3,3.6)
\psline(2.7,3.6)(3.3,3.6)
\psline(7.7,2.6)(8.3,2.6)
\psline(9.7,2.6)(10.3,2.6)
\psline(0.7,1.6)(1.3,1.6)
\psline(2.7,1.6)(3.3,1.6)
\psline(3.7,.6)(4.3,.6)
\psline(5.7,.6)(6.3,.6)

}

{\psset{dotstyle=o,xunit=10pt,yunit=10pt}
\psdots(1,6)(3,6)(5,6)(7,6)(10,6)
\psdots(1,5)(3,5)(6,5)(8,5)(10,5)
\psdots(2,4)(4,4)(6,4)(8,4)(10,4)
\psdots(1,3)(3,3)(6,3)(9,3)(11,3)
\psdots(2,2)(4,2)(6,2)(9,2)(11,2)
\psdots(2,1)(5,1)(7,1)(9,1)(11,1)

\rput(13,6){{\scriptsize : 0}}
\rput(13,5){{\scriptsize : 1}}
\rput(13,4){{\scriptsize : 2}}
\rput(13,3){{\scriptsize : 2}}
\rput(13,2){{\scriptsize : 3}}
\rput(13,1){{\scriptsize : 4}}
}

\end{pspicture}
\end{center}
\end{figure}



Let us now turn to the analysis of the more general situation where a $j$-block is inserted within a $\ell$-block. The insertions within the substring ${b^*}^\ell$ lead to configurations of the form:
\begin{equation}\label{inse}
{b^*}^{i} \,  (b^{*j}\, {b}^j)\,  {b^*}^{\ell-i}\,  b^\ell.
\end{equation}
The partial charge $q_{2\ell-i+j}$ (computed as usual from right to left) is $j+i$; in order to respect the bound $0\leq q_n\leq \ell$, we need to have $i+j\leq \ell$. 
This excludes the insertion at the center of the $\ell$-block ($i=\ell$) and its vicinity.  There are thus $\ell-j+1$ distinct insertions within the substring ${b^*}^\ell$, counting the one with $i=0$ (which is the original configuration where the $j$-block follows the $\ell$-block). Within the substring ${b}^\ell$, there are also $\ell-j+1$ possible insertions, but one of these is identical to one already considered in (\ref{inse}) and it should not be counted twice. The total number of distinct cases is $2(\ell-j)+1$. Each unit displacement of the $j$-block within the $\ell$-block  increases the weight by 1. The counting of possibilities, taking the weight change into account, is thus 
\begin{equation} \begin{bmatrix}
1+2(\ell-j)\\ 1\end{bmatrix}.
\end{equation} 
 This combinatorial factor and the weight shift are nicely exemplified by  our pictorial representation. 
  In the case where a $j$-block is inserted within a $\ell$-block,  we displace the leftmost pair of dots that are separated by $j$ units ($j-1$ dots and 1 hole) and which are free to move simultaneously without generating a sequence of $\ell+1$ successive dots. The displacements are toward the left, by one unit each time. For  $j=2$ and $\ell=4$, this leads to the configurations shown in Fig. \ref{figC}.
 The moved dots are underlined. To reach the fourth configuration, we observe that the leftmost  holes cannot be filled due to the charge constraint. In the operator language, this means that we need to switch from the resulting configuration to its symmetrical counterpart and start the displacements from there on. In other words, in our pictorial representation, the next move involves the filling of the next two holes.


\begin{figure}[ht]
\caption{The five configurations associated to the mixing of a 2-block into a 4-block.}
\vskip-0.6cm
\label{figC}
\begin{center}
\begin{pspicture}(0,0)(5.0,3.0)

{\psset{dotstyle=*,xunit=10pt,yunit=10pt}
\psdots(0,5)(1,5)(3,5)(4,5)(6,5)(7,5)(8,5)(9,5)(11,5)(12,5)(13,5)(14,5)
\psdots(0,4)(1,4)(2,4)(4,4)(5,4)(7,4)(8,4)(9,4)(11,4)(12,4)(13,4)(14,4)
\psdots(0,3)(1,3)(2,3)(3,3)(5,3)(6,3)(8,3)(9,3)(11,3)(12,3)(13,3)(14,3)
\psdots(0,2)(1,2)(2,2)(3,2)(5,2)(6,2)(7,2)(9,2)(10,2)(12,2)(13,2)(14,2)
\psdots(0,1)(1,1)(2,1)(3,1)(5,1)(6,1)(7,1)(8,1)(10,1)(11,1)(13,1)(14,1)
\psline(1.7,3.6)(2.3,3.6)
\psline(4.7,3.6)(5.3,3.6)
\psline(2.7,2.6)(3.3,2.6)
\psline(5.7,2.6)(6.3,2.6)
\psline(6.7,1.6)(7.3,1.6)
\psline(9.7,1.6)(10.3,1.6)
\psline(7.7,.6)(8.3,.6)
\psline(10.7,.6)(11.3,.6)

\rput(15,5){{\scriptsize : 0}}
\rput(15,4){{\scriptsize : 1}}
\rput(15,3){{\scriptsize : 2}}
\rput(15,2){{\scriptsize : 3}}
\rput(15,1){{\scriptsize : 4}}
}

{\psset{dotstyle=o,xunit=10pt,yunit=10pt}
\psdots(2,5)(5,5)(10,5)
\psdots(3,4)(6,4)(10,4)
\psdots(4,3)(7,3)(10,3)
\psdots(4,2)(8,2)(11,2)
\psdots(4,1)(9,1)(12,1)
}

\end{pspicture}
\end{center}
\end{figure}
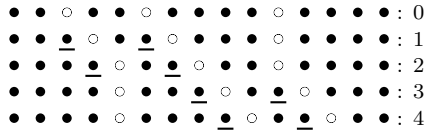



 The description of the various distinct configurations obtained from mixing the blocks in terms of the successive displacements (of one unit toward the left) of two dots separated by $j$ units  readily implies that the  weight  increases by one at each step.
 


The combinatorial factor  
with $p_j$ $j$-blocks and $p_\ell$ $\ell$-blocks is the $q$-deformation of the binomial coefficient that counts the number of ways the $p_j$ $j$-blocks can be inserted within the larger ones. This is 
\begin{equation} \begin{bmatrix}
p_j+2(\ell-j)p_\ell\\ p_j\end{bmatrix}.
\end{equation}

The generalization to a generic sequence of blocks is 
\begin{equation}\label{mix}
\prod_{j=1}^{k-2} \begin{bmatrix}
p_j+2p_{j+1}+4p_{j+2}+\cdots +2(k-1-j)p_{k-1}\\ p_j\end{bmatrix} \equiv  \prod_{j=1}^{k-2} \begin{bmatrix}
p_j+\p_j \\ p_j\end{bmatrix}.
\end{equation}
with
\begin{equation}\label{defpt}
\p_j= 2p_{j+1}+4p_{j+2}+\cdots +2(k-1-j)p_{k-1}.
\end{equation}
Note that if we extend the notation $\p_j$ to the case where the index is 0, we have
\begin{equation}\p_0= 2p_{1}+4p_{2}+\cdots +2(k-1)p_{k-1}=P,
\end{equation}
where $P$ is the number of operator defined in (\ref{defP}).

\subsection{The generating function}

We have constructed all the pieces that compose the 
 generating function for the string of operators subject to the rules (\ref{basis}).  We now put them together. 
 The generating function  for all the possible displacements of the minimal-weight configuration, with a given block content and   the ordering of the operators maintained fixed, is
 \begin{equation}q^{ pB'p}\, \begin{bmatrix}
p_0+\p_0 \\ p_0\end{bmatrix},
\end{equation}
where
\begin{equation}
pB'p = \sum_{i,j=1}^{k-1}p_iB_{i+1,j+1}p_j.
\end{equation}
We next have to consider  all possible mixing of operators that preserve the block content. With the weight evaluated relative to that of the minimal-weight configuration, this amounts to add the factor (\ref{mix}). The resulting expression is (note the lower bound in the product)
\begin{equation}q^{ pB'p} \, \prod_{j=0}^{k-2} \begin{bmatrix}
p_j+\p_j \\ p_j\end{bmatrix},
\end{equation} 
with $\p_j$ defined in (\ref{defpt}). 
This is the generating function for all strings of operators with given values of $p_1,\cdots , p_{k-1}$. The full generating function, for fixed length, is obtained by summing over all  $p_1,\cdots , p_{k-1}$ compatible with (\ref{defL}):
 \begin{equation}
 G^{(L)}(q)= \sum_{p_1,\cdots ,p_{k-1}\geq 0} q^{ pB'p}\, \prod_{j=0}^{k-2} \begin{bmatrix}
p_j+\p_j \\ p_j\end{bmatrix}.
\end{equation}
This is the finitized vacuum character for the $\M(k+1,k+2)$ minimal model. Notice that the finitization is encoded in the $q$-binomial with $j=0$. With $p_j=n_{j+1}$ and $\p_j=m_{j+1}$, we recover the expression (\ref{chi11}).
 
 The infinite length limit is taken by setting $p_0\rw\y$. For this we use  
\begin{equation} 
\lim_{n\rw\y}  \begin{bmatrix}
n\\ m \end{bmatrix}= \frac{1}{(q)_m}\;,
\end{equation} so that 
\begin{equation}
 G^{(\y)}(q)= \chi_{1,1}(q)= \sum_{p_1,\cdots ,p_{k-1}\geq 0} \frac{ q^{ pB'p} } {(q)_{\p_0}} \, \prod_{j=1}^{k-2} \begin{bmatrix}
p_j+\p_j \\ p_j\end{bmatrix}.
\end{equation}

\subsection{Modules other than the vacuum one}

As already said, the various irreducible modules labeled by the pair $(r,s)$ are 
characterized by their boundary conditions (\ref{rs}) (assuming a final SE edge). Note that the module $(r,s)$ is identified with the one labeled by $(k+1-r,k+2-s)$. Using this equivalence, it is always possible to chose $s\leq r$, that is, $y_0\leq y_L$. The ground state in the case $r=s$ is similar to the one for the vacuum module: it zigzags between the height $s-1$ and $s$.  From this ground state, one can reach the ground state for all the modules with $r>s$ by acting with the string $b_{r-s}\cdots b_{1}$. The complete set of states in the finitized $(r,s)$ module with $r\geq s$  is obtained by adjoining to this initial string a charge-less sequence satisfying the conditions (\ref{basis}) except that the bound on $i_1$ is modified to read $i_1\geq (r-s+1)$ and the partial charge of the added sequence will now satisfy $-(r-s)\leq q_\ell\leq k-1-r+s.$ That different modules $(r\geq s)$ can be described in terms of a single one $(r=s)$ testifies a sort of economy  of the operator description. However, the weights  of the modules so related cannot be read off directly from this description: $h_{r,s}\not = h_{s,s}+1/2+\cdots +(r-s)/2$. (The exception is the Ising model when $s=1$.) We will not detail  the combinatorial analysis of this more general case as it does not significantly differs from that already considered and that no new formulae are so generated.




\section{Operator description of the $\M(k+1,2k+3)$ models}

\subsection{Paths and operators for the $\M(k+1,2k+3)$ models}

It is quite remarkable that the $\M(k+1,2k+3)$ models do have a path description very similar to that of the  $\M(k+1,k+2)$ models \cite{JSTAT}. The essential difference is that the
 lattice is half-integer and peaks (whose charge can range from 1/2 to $k-1/2$) are forced to be at integer $(x,y)$ positions. Here again the paths are defined within the rectangle $0\leq y\leq k$ and $0\leq x\leq L$ (but the lattice delimitated by this rectangle is now half-integer). Moreover, the weight function is defined exactly as for the unitary models: all vertices but the  extrema contribute to $x/2$. The paths start at integer height $y_0$  and end up at half-integer height $y_L$, with a SE edge (where here an edge connects the points $(x,y)$ and $(x+1/2, y\pm 1/2)$). The module labels are related to the boundary values as
 \begin{equation}
 r=y_L+\frac12 \qquad \text{and} \qquad s=2y_0+1.
 \end{equation}

The ground state pertaining to the vacuum module of all the models of this class is displayed in Fig. \ref{fig3}. Observe that it is necessary to have a peak of charge one preceding those of charge 1/2.  An example of path, appropriate to any  model with $k\geq 2$, is displayed in Fig. \ref{fig4}.

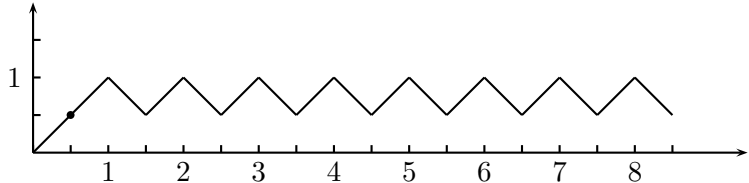
\begin{figure}[ht]
\caption{{\footnotesize The ground-state path in the vacuum module of all $\M(k+1,2k+3)$ models. The single vertex contributing to the weight, which is $1/4$,  is indicated by a dot.}} \label{fig3}
\begin{center}
\begin{pspicture}(0,0)(11.5,3)
\psline{->}(0.5,0.5)(0.5,2.5) \psline{->}(0.5,0.5)(10.0,0.5)
\psset{linestyle=solid}
\psline{-}(0.5,0.5)(0.5,0.6) \psline{-}(1.0,0.5)(1.0,0.6)
\psline{-}(1.5,0.5)(1.5,0.6) \psline{-}(2.0,0.5)(2.0,0.6)
\psline{-}(2.5,0.5)(2.5,0.6) \psline{-}(3.0,0.5)(3.0,0.6)
\psline{-}(3.5,0.5)(3.5,0.6) \psline{-}(4.0,0.5)(4.0,0.6)
\psline{-}(4.5,0.5)(4.5,0.6) \psline{-}(5.0,0.5)(5.0,0.6)
\psline{-}(5.5,0.5)(5.5,0.6) \psline{-}(6.0,0.5)(6.0,0.6)
\psline{-}(6.5,0.5)(6.5,0.6) \psline{-}(7.0,0.5)(7.0,0.6)
\psline{-}(7.5,0.5)(7.5,0.6) \psline{-}(8.0,0.5)(8.0,0.6)
\psline{-}(8.5,0.5)(8.5,0.6) \psline{-}(9.0,0.5)(9.0,0.6)
\rput(1.5,0.25){{\small $1$}}\rput(2.5,0.25){{\small
$2$}}\rput(3.5,0.25){{\small $3$}}\rput(4.5,0.25){{\small
$4$}}\rput(5.5,0.25){{\small $5$}}\rput(6.5,0.25){{\small
$6$}}\rput(7.5,0.25){{\small $7$}}\rput(8.5,0.25){{\small $8$}}

\psline{-}(0.5,1.0)(0.6,1.0) \psline{-}(0.5,1.5)(0.6,1.5)
\psline{-}(0.5,2.0)(0.6,2.0) \rput(0.25,1.5){{\small $1$}}
\psline{-}(0.5,0.5)(1.0,1.0) \psline{-}(1.0,1.0)(1.5,1.5)
\psline{-}(1.5,1.5)(2.0,1.0) \psline{-}(2.0,1.0)(2.5,1.5)
\psline{-}(2.5,1.5)(3.0,1.0) \psline{-}(3.0,1.0)(3.5,1.5)
\psline{-}(3.5,1.5)(4.0,1.0) \psline{-}(4.0,1.0)(4.5,1.5)
\psline{-}(4.5,1.5)(5.0,1.0) \psline{-}(5.0,1.0)(5.5,1.5)
\psline{-}(5.5,1.5)(6.0,1.0) \psline{-}(6.0,1.0)(6.5,1.5)
\psline{-}(6.5,1.5)(7.0,1.0) \psline{-}(7.0,1.0)(7.5,1.5)
\psline{-}(7.5,1.5)(8.0,1.0) \psline{-}(8.0,1.0)(8.5,1.5)
\psline{-}(8.5,1.5)(9.0,1.0)
\psset{dotsize=3pt} \psdots(1.0,1.0)

\end{pspicture}
\end{center}
\end{figure}

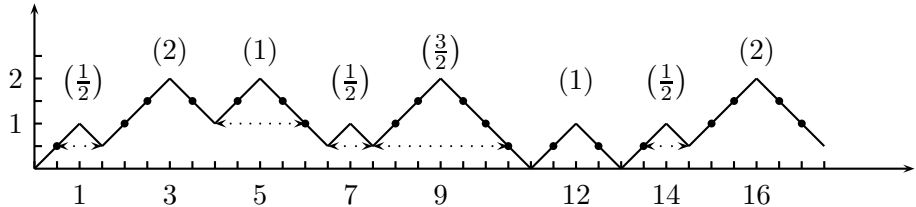
\begin{figure}[ht]
\caption{{\footnotesize An example of a $\M(k+1,2k+3)$ path for $k\geq 2$ starting at $y_0=0$ and ending at $y_{35/2}=1/2$. Weight contributing vertices are indicated by dots. Note that the  $(x,y)$ coordinates of the peaks are integers.  The charge of the  peaks is determined as previously but with the path augmented by a final SE edge. The charge  is  the height measured from the indicated dotted line or the horizontal axis and it is indicated above each peak.}}\label{fig4}
\begin{center}
\begin{pspicture}(0,0)(12.5,3)
\psline{->}(0.3,0.3)(0.3,2.5) \psline{->}(0.3,0.3)(12.0,0.3)
\psset{linestyle=dotted} \psline{<->}(0.6,0.6)(1.2,0.6)
\psline{<->}(4.2,0.6)(4.8,0.6) \psline{<->}(4.8,0.6)(6.6,0.6)
\psline{<->}(2.7,0.9)(3.9,0.9) \psline{<->}(8.4,0.6)(9.0,0.6)
\psset{linestyle=solid}
\psline{-}(0.3,0.3)(0.3,0.4) \psline{-}(0.6,0.3)(0.6,0.4)
\psline{-}(0.9,0.3)(0.9,0.4) \psline{-}(1.2,0.3)(1.2,0.4)
\psline{-}(1.5,0.3)(1.5,0.4) \psline{-}(1.8,0.3)(1.8,0.4)
\psline{-}(2.1,0.3)(2.1,0.4) \psline{-}(2.4,0.3)(2.4,0.4)
\psline{-}(2.7,0.3)(2.7,0.4) \psline{-}(3.0,0.3)(3.0,0.4)
\psline{-}(3.3,0.3)(3.3,0.4) \psline{-}(3.6,0.3)(3.6,0.4)
\psline{-}(3.9,0.3)(3.9,0.4) \psline{-}(4.2,0.3)(4.2,0.4)
\psline{-}(4.5,0.3)(4.5,0.4) \psline{-}(4.8,0.3)(4.8,0.4)
\psline{-}(5.1,0.3)(5.1,0.4) \psline{-}(5.4,0.3)(5.4,0.4)
\psline{-}(5.7,0.3)(5.7,0.4) \psline{-}(6.0,0.3)(6.0,0.4)
\psline{-}(6.3,0.3)(6.3,0.4) \psline{-}(6.6,0.3)(6.6,0.4)
\psline{-}(6.9,0.3)(6.9,0.4) \psline{-}(7.2,0.3)(7.2,0.4)
\psline{-}(7.5,0.3)(7.5,0.4) \psline{-}(7.8,0.3)(7.8,0.4)
\psline{-}(8.1,0.3)(8.1,0.4) \psline{-}(8.4,0.3)(8.4,0.4)
\psline{-}(8.7,0.3)(8.7,0.4) \psline{-}(9.0,0.3)(9.0,0.4)
\psline{-}(9.3,0.3)(9.3,0.4) \psline{-}(9.6,0.3)(9.6,0.4)
\psline{-}(9.9,0.3)(9.9,0.4) \psline{-}(10.2,0.3)(10.2,0.4)
\psline{-}(10.5,0.3)(10.5,0.4) \psline{-}(10.8,0.3)(10.8,0.4)

\rput(0.9,-0.05){{\small $1$}} \rput(2.1,-0.05){{\small $3$}}
\rput(3.3,-0.05){{\small $5$}} \rput(4.5,-0.05){{\small
$7$}}\rput(5.7,-0.05){{\small $9$}} \rput(7.5,-0.05){{\small $12$}}
\rput(8.7,-0.05){{\small $14$}}\rput(9.9,-0.05){{\small $16$}}
 \psline{-}(0.3,0.6)(0.4,0.6)
\psline{-}(0.3,0.9)(0.4,0.9) \psline{-}(0.3,1.2)(0.4,1.2)
\psline{-}(0.3,1.5)(0.4,1.5) \psline{-}(0.3,1.8)(0.4,1.8)

\rput(0.05,0.9){{\small $1$}} \rput(0.05,1.5){{\small $2$}}
\psline{-}(0.3,0.3)(0.6,0.6) \psline{-}(0.6,0.6)(0.9,0.9)

\psline{-}(0.9,0.9)(1.2,0.6)

\psline{-}(1.2,0.6)(1.5,0.9) \psline{-}(1.5,0.9)(1.8,1.2)
\psline{-}(1.8,1.2)(2.1,1.5)

\psline{-}(2.1,1.5)(2.4,1.2) \psline{-}(2.4,1.2)(2.7,0.9)

\psline{-}(2.7,0.9)(3.0,1.2) \psline{-}(3.0,1.2)(3.3,1.5)

\psline{-}(3.3,1.5)(3.6,1.2) \psline{-}(3.6,1.2)(3.9,0.9)
\psline{-}(3.9,0.9)(4.2,0.6)

\psline{-}(4.2,0.6)(4.5,0.9)

\psline{-}(4.5,0.9)(4.8,0.6)

\psline{-}(4.8,0.6)(5.1,0.9) \psline{-}(5.1,0.9)(5.4,1.2)
\psline{-}(5.4,1.2)(5.7,1.5)

\psline{-}(5.7,1.5)(6.0,1.2)\psline{-}(6.0,1.2)(6.3,0.9)
\psline{-}(6.3,0.9)(6.6,0.6) \psline{-}(6.6,0.6)(6.9,0.3)

\psline{-}(6.9,0.3)(7.2,0.6) \psline{-}(7.2,0.6)(7.5,0.9)

\psline{-}(7.5,0.9)(7.8,0.6) \psline{-}(7.8,0.6)(8.1,0.3)

\psline{-}(8.1,0.3)(8.4,0.6) \psline{-}(8.4,0.6)(8.7,0.9)

\psline{-}(8.7,0.9)(9.0,0.6)

\psline{-}(9.0,0.6)(9.3,0.9) \psline{-}(9.3,0.9)(9.6,1.2)
\psline{-}(9.6,1.2)(9.9,1.5)

\psline{-}(9.9,1.5)(10.2,1.2) \psline{-}(10.2,1.2)(10.5,0.9)
\psline{-}(10.5,0.9)(10.8,0.6)
\psset{dotsize=3pt}
\psdots(0.6,0.6)(1.5,0.9)(1.8,1.2)(2.4,1.2)(3.0,1.2)(3.6,1.2)
(3.9,0.9)(5.1,0.9)(5.4,1.2)(6.0,1.2)(6.3,0.9)(6.6,0.6)(7.2,0.6)
(7.8,0.6)(8.4,0.6)(9.3,0.9)(9.6,1.2)(10.2,1.2)(10.5,0.9)

\rput(0.95,1.45){{\small $\left(\frac12\right)$}} 
\rput(2.1,1.85){{\small $(2)$}}
\rput(3.3,1.85){{\small $(1)$}} 
\rput(4.5,1.45){{\small $\left(\frac12\right)$}}
\rput(5.7,1.85){{\small $\left(\frac32\right)$}} 
\rput(7.5,1.45){{\small $(1)$}}
\rput(8.7,1.45){{\small $\left(\frac12\right)$}}
\rput(9.9,1.85){{\small $(2)$}}

\end{pspicture}
\end{center}
\end{figure}

Like for the unitary case, the path can be dressed with a natural operator interpretation, in terms of the operators $b$ and $b^*$ defined as before (although their modes are now multiple of 1/2). Note however that the ground-state path cannot be regarded as resulting from the action of $b_{1/2}$ on the path that would zigzag between 0 and 1/2 since such a path is not allowed, its peaks lying at half-integer positions. Therefore, we consider this $b_{1/2}$ to be built-in the vacuum definition. The partial charge (which is again computed from right to left) in this case satisfies
 \begin{equation}\label{qn}
-1\leq q_\ell\leq 2(k-1).
 \end{equation}
For instance, to the path of Fig. \ref{fig4}, we associate the sequence:  
$b^*_{34/2}b^*_{33/2}b_{31/2}b_{30/2}b_{27/2}\cdots b_{4/2}$.


Our task  is again to write down  the basis pertaining to these models. But to motivate some of the differences with the $\M(k+1,k+2)$ basis, it is convenient to consider briefly the first  model $(k=1)$ of the sequence, focusing again on the simple vacuum boundary conditions $(y_0,y_L)=(0,1/2)$. Appropriate states  are thereby described by charge-less sequences of operators.
 
 \subsection{The $\M(2,5)$ model}
 
 For $k=1$,  the only allowed operator combinations
 are of type $bb^*$. Moreover, these pairs must necessarily occur in closely-packed doublets, in the form $ b_{i+1}b^*_i$ because they cannot be separated by peaks of charge $1/2$ since these would then necessarily create peaks at non-integer positions.

Let $p$ be the number of 1-block of the type $bb^*$. The minimal-weight configuration with $p$ such pairs is $ b_{3/2+2p-1} \cdots b^*_{7/2}b_{5/2}b^*_{3/2}$ and its weight is easily found to be $p^2+p$. Displacements need to be considered but each displacement must involve the two members of the block.  Both mode indices change then by steps of $1$,  so that the weight change for every block displacement is $(1+1)/2=1$. The maximal displacement of the leftmost $b$ is  from $3/2+2p-1$ to $L-1$ (this last possible position being determined by the constraint of a final SE edge). The maximal displacement is thus $L-3/2-2p$ and since the two parts of the doublet are moved together, this is the weight shift. Since there are $p$ blocks, one has to $q$-enumerate the number of partitions into at most $p$ parts with the maximal part $\leq  L-3/2-2p$. This is given by the combinatorial factor
\begin{equation} 
 \begin{bmatrix}
L-2p-3/2+p\\ p \end{bmatrix} = \begin{bmatrix}
L-p-3/2\\ p \end{bmatrix}.
\end{equation} 
 The generating function reads
\begin{equation} 
 \sum_{p\geq 0}q^{p^2+p} \begin{bmatrix}
L-2p-3/2\\ p \end{bmatrix}\quad \xrightarrow{L\rw\y}\quad \sum_{p\geq 0}\frac{ q^{p^2+p}}{(q)_p}.
\end{equation}  
This is the expected result. To compare with the expressions presented in 
 \cite{JSTAT}, note that $p=n_1-1$, where $n_1$ is the number of peaks of charge 1.
 
 
 \subsection{The $\M(k+1,2k+3)$ operator basis}
 
For the $\M(3,7)$ model, the only allowed operator combinations
 are of type $bb^{*3}b^2$, $b^{*2}b^{2}$ or $bb^*$, which we will call 3-, 2- and 1-blocks respectively.
 In the general case, one needs to consider $\ell$-blocks with $1\leq \ell \leq 2k-1$, with the following structure:
 \begin{align} 
 &bb^{*\ell}b^{\ell-1}\quad \text{for $\ell$ odd}\nonumber\\
& 
b^{*\ell} b^{\ell}\quad\;\,\quad\text{for $\ell$ even}.
 \end{align}   
 The strategy for constructing the generating function is again to identify the minimal-weight configuration for a fixed block content,   determine its weight, $q$-enumerate all possible configurations, and sum over the block content. 

The problem of main interest in this novel context 
is to unravel the conditions to be imposed on successive operators which  ensure  the peaks, in the associated path representation,  to lie at integer positions. 
As a first orientation on this question, consider the insertion of a 1-block within a 3-block. 
%
The allowed  possibilities are the following:
 \begin{equation}
 (bb^*)bb^*b^*b^*bb, \quad
bb^*(bb^*)b^*b^*bb, \quad
bb^*b^*b^*(bb^*)bb, \quad
bb^*b^*b^*bb(bb^*), 
\end{equation}
(i.e., the 1-block is moved by two units in each step) while those that must be discarded are
 \begin{equation}
 b(bb^*)b^*b^*b^*bb, \quad
bb^*b^*(bb^*)b^*bb, \quad
 bb^*b^*b^*b(bb^*)b.
\end{equation} 
The first discarded one does not satisfies the bound (\ref{qn}) (i.e., $q_6=-2$). The other two are not allowed because they have a peak at half-integer position.

It is clear from this example that to ensure the integrality of the peak position, which occurs in-between a $b$ and a $b^*$,  one must impose a condition formulated in terms of the parity of the partial charge up to and including  the operator $b$ just before the $b^*$ to be inserted.  This condition is simply that the partial charge before the insertion of $b^*$ needs to be even:
\begin{equation} \label{pch}\cdots b^*_{i_{\ell+1}}  b_{i_\ell}\cdots : \quad q_\ell =2n.
\end{equation} 
Again we stress that the computation of the partial charge must not take into account the operator $b_{1/2}$ which is always present and considered to be part of the mere definition of the vacuum.

Finally, there is also a potential constraint on the modes of adjacent operators that generalizes the closely-packed condition of the pairs  $bb^*$ 
in the $\M(2,5)$ model. With $c$ being either $b$ or $b^*$, the generalized form of this condition is the following: to the usual hard-core repulsion 
\begin{equation} \label{pcc}\cdots  c_{i_{\ell+1}} c_{i_\ell}\cdots :  \quad i_{\ell+1}-i_\ell
=n+\frac12+\frac14[1-q(c_{i_\ell})q(c_{i_{\ell+1}})]
\end{equation} 
one adds the condition
\begin{equation}\label{pco} n=0 \quad  \text{if}\quad  q_\ell =2m+1.
\end{equation}  
 
These  conditions (\ref{pch}) (\ref{pcc}) and (\ref{pco})
%
 are the essential new characteristics defining the basis as compared to the unitary one.

This completes our presentation of the operator  basis for the  $\M(k+1,2k+3)$ models. The combinatorial analysis of these models, in terms of paths, is presented in \cite{JSTAT} and it will not be rephrased in the operator basis. Again, the resulting character formulae are the same.

\section{Concluding remarks}

We have presented a rather natural operator description of the states in the irreducible modules of the unitary minimal models. This fermionic-type operator construction is derived from the RSOS path representation of the states. On the paths, these operators act in a non-local way. This construction can be abstracted and made independent of the path used for their definition. This leads to the formulation of the set of conditions (\ref{basis}) that define a basis. The correctness of these basis conditions  has been supported by the derivation (relying solely on (\ref{basis})) of the generating function that reproduce the known fermionic characters \cite{OleJS}. In this abstract form, an element of non-locality is still present through a condition involving the partial charge of a sequence of operators.

The operational description is somewhat economical when compared to the path in that we do not have to take into account peaks of charge 1
which do not contribute to the weight. This  is similar to the relation between Bressoud  and RSOS paths in the description of the $\z_k$  parafermions \cite{Path}: the variables in the former case are peaks of charge between 1 and $k-1$ while  peaks of charge 1 up  to $k$  are  considered in the RSOS case. Although peaks of charge $k$ do not contribute to the weight, they allow for a natural way of finitizing the paths. In contrast, the Bressoud paths are not naturally finitized.
 Similarly, the minimal models have a natural finitization in their  RSOS path formulation, while the operator approach does not entail such a natural finite form.

Let us expand a bit on the issue of finitization. The way we have finitized the characters in the operator basis by constraining the maximal value of the rightmost operator is of course dictated by the path representation that constitute the starting point of the construction. But once the basis is written and taken from an intrinsic standpoint, outside its original path context, the consideration of a finite length becomes rather artificial. It is thus more natural to set $L\rw \y$ in (\ref{basis}). The generating function of the resulting basis elements leads then directly to the  conformal character. In this regard, it is appropriate to recall one of  the main reason that makes  finitized expressions  interesting: Finitization offers a powerful way of proving the correctness of conjectured fermionic expressions by demonstrating that they satisfy the same recurrence relations that characterize the finitized configuration sums (see e.g., \cite{Mel,Ber}).
But this motivation becomes much less striking within the context of a constructive method such as the one presented here (and \cite{OleJS}).\footnote{{Apart from their intrinsic interest, the other motivation for finitized characters has already pointed out in the introduction, which is the possibility of defining dual characters from the $q\rw q^{-1}$ transformation \cite{BMlmp,OleJS, Kyoto,FWa}.}}

When considered from the point of view of the general Forrester-Baxter RSOS paths, the operator method presented here applies only to the unitary minimal models. The crucial simplifying feature of the unitary models is that the weight of the contributing vertices (those with non-zero weight) depends solely upon their $x$-position.
However, we have shown in \cite{JSTAT} that the $\M(k+1,k+2)$ Bailey duals \cite{FLPW}, the  $\M(k+1,2k+3)$ models, do have a similar  path description (albeit without yet a RSOS underlying construction). It it thus not surprising to find that our operator setting can be lifted to that case.



But what is probably more compelling (and which, ultimately, might be viewed as the best immediate argument justifying the interest of this new approach) is that this operator construction can be extended to the superconformal unitary models. For these models, 
the constructive method following the lines of \cite{OleJS} is not  so directly worked out. It turns out that in this context,  the logic underlying the fermi-gas description of the paths is reversed: a path has a natural  operator construction and it is this very operator  interpretation which, once finitized, allows us to clearly identify the particles within the path  \cite{JMsusy}. We suspect that the same situation will hold for the fermi-gas analysis of the more general models \cite{Kyoto}
$\widehat {su}(2)_k\otimes \widehat {su}(2)_\ell /\widehat {su}(2)_{k+\ell}$, for $\ell>2$: working at the level of the operator basis induced by the path description is likely to be simpler than working directly at the level of the paths.

There are a number of natural extensions of this work. At first, there is a major loose end from the point of view of our motivating vague discussion in the introduction, which led us to suspect the operator basis to be non-local and potentially related to  parafermionic-type quasi-particles. Clearly, the whole conformal-field-theoretical interpretation of these operators still has to be unravelled.
Finally, it would be of interest  to see if this new operator could be used to understand the action of the Virasoro algebra on the paths, in the spirit of previous works (see e.g. \cite{RV} and references therein).



  \end{document}